\documentclass[twocolumn,secnumarabic,amssymb, nobibnotes, aps, prd]{revtex4}
\usepackage{graphicx}

\begin{document}
\title{A problem with the conservation law observed in macroscopic quantum phenomena is a consequence of violation of the correspondence principle.}
\author{A.V. Nikulov}
\email[]{nikulov@iptm.ru}
\affiliation{Institute of Microelectronics Technology and High Purity Materials, Russian Academy of Sciences, 142432 Chernogolovka, Moscow District, RUSSIA.} 
\begin{abstract} Jorge Hirsch drew attention to the contradiction of the Meissner effect with Faraday's law and the law of angular momentum conservation. This article draws attention to the fact that the angular momentum can change without any force only due to quantization in both microscopic and macroscopic quantum phenomena. But if in the first case this change cannot exceed the Planck constant, then in the second case this change is macroscopic due to a violation of the correspondence principle. To explain the violation of the correspondence principle, Lev Landau postulated in 1941 that microscopic particles in superfluid helium and superconductor cannot move separately. Quantization can change not only the angular momentum, but also the kinetic energy of a superconducting condensate on a macroscopic amount. For this reason, the Meissner effect and other macroscopic quantum phenomena contradict the second law of thermodynamics. The reluctance of physicists to admit the violation of the second law of thermodynamics provoked a false understanding of the phenomenon of superconductivity and obvious contradictions in books on superconductivity.
\end{abstract}

\maketitle 

\narrowtext

\section{Introduction}
\label{}
The theory of superconductivity, proposed in 1957 by J. Bardeen, L.N. Cooper and J.R. Schrieffer (BCS), is considered an outstanding achievement of theoretical physics of the 20th century. But Jorge Hirsch tries to convince the community of superconductivity researchers in his numerous articles and in the book \cite{Hirsch2020book} that the theory of BCS \cite{BCS1957} is internally contradictory and cannot describe some macroscopic quantum phenomena observed in superconductors. Hirsch writes in the book \cite{Hirsch2020book} that his point of view is considered heresy in the scientific world, since the theory of BCS is universally considered correct. The history of science knows many examples when what was considered heresy eventually gained universal acceptance. Hirsch's arguments against BCS theory should be considered critically, not ignored. This work reveals the causes of the puzzle and contradictions that Hirsch pointed out.

One of Hirsch's main arguments against the BCS theory is the inability of this theory to solve the puzzle \cite{Hirsch10Meis} of the Meissner effect \cite{Meissner1933}. Hirsch draws attention to the fact that "{\it within the conventional understanding of superconductivity electrons appear to change their velocity in the absence of Lorentz forces}" \cite{Hirsch2003Lorentz}. The conventional theory of superconductivity cannot eliminate the obvious contradiction of the Meissner effect with Lenz's law and the law of angular momentum conservation \cite{Hirsch2007Lenz}. Hirsch argues that the London-BCS theory cannot explain magnetic flux expulsion from both a bulk superconductor \cite{Hirsch10Meis} and in a superconducting wire \cite{Hirsch2021wire}. 

The essence of the Meissner effect puzzle is that the shielding superconducting currents (which by W. H. Keesom called persistent currents \cite{Keesom1934}) emerge in the absence of a known force. Hirsch expressed surprise: "{\it Strangely, the question of what is the 'force' propelling the mobile charge carriers and the ions in the superconductor to move in direction opposite to the electromagnetic force in the Meissner effect was essentially never raised nor answered to my knowledge, except for the following instances: \cite{LondonH1935} (H. London states: 'The generation of current in the part which becomes supraconductive takes place without any assistance of an electric field and is only due to forces which come from the decrease of the free energy caused by the phase transformation,' but does not discuss the nature of these forces), \cite{PRB2001} (A.V. Nikulov introduces a 'quantum force' to explain the Little-Parks effect in superconductors and proposes that it also explains the Meissner effect)}" \cite{Hirsch10Meis}. 

In the article \cite{HirschAPS} Hirsch compares the attitude of most experts to this puzzle with the attitude of the characters of the fairy tale "The Emperor's New Clothes" by Hans Christian Andersen to the emperor's new clothes: 

"{\it '"Heaven help me" he thought as his eyes flew wide open, 'I can't see anything at all'. But he did not say so.'

'Heaven help me', thought smart students that couldn't understand how BCS theory explains the Meissner effect. 'I can't possibly see how momentum conservation is accounted for and Faraday's law is not violated'. But they did not say so}" \cite{HirschAPS}. 

The puzzle of the conservation laws noticed by Hirsch and the smart students in the Meissner effect is not new in quantum mechanics. Bohr and his co-authors expressed doubts about the conservation laws in elementary quantum processes in an article \cite{Bohr1924} published in 1924. Later, Bohr came to the conclusion that the uncertainty relation and the rejection of an accurate description of space-time processes can eliminate the contradiction between the quantum theory and the conservation laws \cite{Bohr1936}. But the uncertainty relation can eliminate this contradiction, unless the change in angular momentum due to quantization exceeds Planck's constant. While the change in angular momentum by a macroscopic amount significantly exceeding Planck's constant is observed in the Meissner effect and other quantum phenomena observed in superconductors. 

Quantum phenomena were first observed at the atomic, microscopic level, and quantum theory was created on the basis of the correspondence principle, according to which quantum-theoretic conclusions converge to classical results at the macroscopic level, at which the quantum of action, i.e. Planck's constant $h$, can be considered an infinitesimal value $h \rightarrow 0$. A quantum theory that does not violate the correspondence principle may be called a microscopic quantum theory. A theory must violate the correspondence principle in order to describe macroscopic quantum phenomena. Such a theory may be called a macroscopic quantum theory. 

The angular momentum can change without any known force, only by quantization in accordance with both microscopic and macroscopic quantum theory. But according to the microscopic quantum theory, i.e. Bohr's quantization and Schrodinger's wave mechanics, this change cannot exceed the reduced Planck constant $\hbar = h/2\pi$ due to the correspondence principle, while according to the macroscopic quantum theory, i.e. the conventional theory of superconductivity \cite{BCS1957,GL1950}, the change in angular momentum due to quantization can be macroscopic. This distinction is very important because of Bohr's remark \cite{Bohr1936} that only the uncertainty relation saves quantum theory from contradicting conservation laws. In this article, the difference between microscopic and macroscopic quantum phenomena is demonstrated by the example of the persistent current of electrons observed in mesoscopic normal metal rings and the persistent current of Cooper pairs observed in superconductor rings. 

The next section reminds readers about the sense and history of the correspondence principle. The persistent current of electrons observed in normal metal rings \cite{PCScien09,PCPRL09} is considered in the third section. The persistent currents of Cooper pairs considered in the fourth section contradict the correspondence principle, but are successfully described by the Ginzburg-Landau theory \cite{GL1950}. A comparison of the change in angular momentum due to quantization in the case of the persistent current of electrons and of Cooper pairs in the fifth section proves that the puzzle with the law of angular momentum conservation is observed due to a violation of the correspondence principle in macroscopic quantum phenomena. The fundamental difference between  the GL wave function and the Schrodinger wave function, due to the Landau postulate, is considered in the sixth section.  
 
Hirsch draws the reader's attention in the articles \cite{HirschEPL,HirschIJMP,HirschPhys} to the fact that the conventional theory of superconductivity \cite{BCS1957,GL1950} contradicts the laws of thermodynamics. The contradiction of this theory \cite{BCS1957,GL1950} and also of macroscopic quantum phenomena to the second law of thermodynamics is considered in the seventh section. In the eighth section, attention is drawn to the contradictions in books on superconductivity, provoked by the unwillingness to recognize the obvious contradiction of the Meissner effect and other macroscopic quantum phenomena to the second law of thermodynamics. In Conclusion, the reasons for the mass misconception are discussed. 

\section{Correspondence Principle}
 \label{}
Hirsch is one of the few superconductivity experts who have noticed the discrepancy between the observation of macroscopic quantum phenomena and the Bohr correspondence principle. He writes about the electric current in a superconducting ring, which persists for years without fading, like electron orbits in atoms: "{\it It also has important consequences due to what is known as 'Bohr's correspondence principle'. In 1913, Niels Bohr relied on the principle that microscopic physics should smoothly evolve into macroscopic physics as the dimensions of the system grow in order to deduce the laws that govern the behavior of the hydrogen atom. Surprisingly, however, this principle is not taken into account in the contemporary science of superconductivity. It relies on microscopic physics and disregards Bohr's correspondence principle"} \cite{Hirsch2020book}. 

Max Jammer wrote about the correspondence principle in his book about the history of quantum mechanics: "{\it The conceptual foundation of the correspondence principle was based on the assumption that quantum theory or at least its formalism contains classical mechanics as a limiting case. This idea was expressed by Planck as early as 1906, when he showed that in the limit $h \rightarrow 0$ quantum-theoretic conclusions converge toward classical results, or, as he said: 'The classical theory can simply be characterized by the fact that the quantum of action becomes infinitesimally small'"} \cite{Jammer1966}. Bohr's correspondence principle, according to Max Jammer, does not differ fundamentally from Planck's idea: "{\it The same result, however, is also obtained if for constant $h$ the frequency $\nu $ approaches zero. This was the idea chosen by Bohr in his establishment of the correspondence principle. For, according to his frequency law, the convergence of $\nu $ toward zero implies that the energy differences become arbitrarily small or, in other words, that in the limit the energies form almost a continuum"} \cite{Jammer1966}. 

The discreteness of the spectrum of permitted states is one of the main differences between quantum mechanics and classical mechanics. The discreteness of the spectrum cannot be noticeable when the energy difference between the permitted states is much less than the thermal energy per particle $\Delta E \ll k_{B}T$. The correspondence principle implies that there should be a transition from a discrete spectrum $\Delta E \gg k_{B}T$ to a continuous one $\Delta E \ll k_{B}T$ with an increase in the size, for example, of the orbit, since the discreteness of the spectrum is proportional to the square of the Planck constant $\Delta E \propto \hbar^{2}$. An experimental example of the intermediate discreteness $\Delta E \approx k_{B}T$, when the magnitude of the quantum effect depends exponentially on size and temperature, will be considered in the next section. 

\section{The persistent currents of electrons satisfy the correspondence principle}
 \label{}  
It is believed \cite{Jammer1966} that the correspondence principle was formulated by Niels Bohr in 1920 \cite{Bohr1920}, although he had used it back in 1913 when developing his model of the atom \cite{Bohr1913}. According to the second postulate of the Bohr model, stationary orbits must correspond to an integer $n$ = 1, 2, 3, ... multiple of the reduced Planck constant $\hbar = h/2\pi$: $pr = mvr = n\hbar$. The Sommerfeld quantization condition \cite{Jammer1966}
$$\oint_{l}dl p = nh = n 2\pi \hbar \eqno{(1)}$$ 
generalized this postulate to electron orbits of any shape. According to the de Broglie relation $p = h/\lambda $ between wavelength of an electron $\lambda $ and its momentum $p = mv$, stationary orbits must correspond to an integer number of the wavelengths: $\oint_{l}dl p = h\oint_{l}dl 1/\lambda = hl/\lambda = hn$. According to the Schrodinger wave mechanics \cite{Schrodinger1926}, the Sommerfeld quantum condition (1) can be considered as a consequence of the requirement that the complex wave function $\Psi = |\Psi |e^{i\varphi }$ must be single-valued  $\Psi = |\Psi |e^{i\varphi } =  |\Psi |e^{i(\varphi + n2\pi )} $ at any point along the integration path $l$: $\oint_{l}dl p = \hbar \oint_{l}dl \nabla \varphi = 2\pi \hbar n $ since the momentum of a quantum particle $p = \hbar \nabla \varphi $ when the value $|\Psi(\vec{r})|$ does not depend on the coordinate $\vec{r}$. This requirement can be applied to both atomic orbits and a ring with a radius $r$.

The differences in velocity $v = p/m = \hbar n/rm$ and kinetic energy 
$$\Delta E = E_{n+1} - E_{n} = \frac{p_{n+1}^{2}}{2m} - \frac{p_{n}^{2}}{2m} = (2n+1)\frac{\hbar ^{2}}{2mr^{2}} \eqno{(2)}$$ 
between adjacent permitted states $n+1$ and $n$ decrease with an increase in the mass $m$ of a particle and the ring radius $r$. Relation (2) proves that Bohr quantization satisfies the correspondence principle: the spectrum (2) is discrete $\Delta E \gg k_{B}T$ at a microscopic radius $r$ and mass $m$, and is continuous $\Delta E \ll k_{B}T$ at a macroscopic radius $r$ or mass $m$. The Schrodinger wave mechanics also satisfies the correspondence principle. The Schrodinger equation converges toward the classical Hamilton-Jacobi equation for the action $S$ of a particle in the Planck limit $h \rightarrow 0$, more exactly when the action of the particle has a macroscopic magnitude $S \gg \hbar $ \cite{LandauL}.  

According to the correspondence principle, the quantum effects associated with Bohr's quantization (1) should not be observed in the case of weak discreteness, when the energy difference (2) between states $n+1$ and $n$ is much less than the thermal energy per particle $\Delta E \ll k_{B}T$. Temperature should not affect quantization in the opposite case of strong discreteness $\Delta E \gg k_{B}T$, which occurs in atoms: the energy difference between adjacent permitted states $\Delta E \approx \hbar ^{2}/2mr^{2} \approx 2 \ 10^{-18} \ J$ for the Bohr radius $r_{B} \approx 0.05 \ nm = 5 \ 10^{-11} \ m$ corresponds to a temperature $T = \Delta E/k _{B} \approx 100000 \ K$ significantly higher than room temperature $T \approx 300 \ K$. 

Quantization effects can be observed and depends on the temperature only at an intermediate discreteness $\Delta E \approx k_{B}T$. The measurement results of a such quantum phenomenon as the persistent current of electrons observed in normal metal nano-rings with radius $r \approx  300 \div 800 \ nm$ \cite{PCScien09,PCPRL09} corroborate this prediction of the correspondence principle. The persistent current is observed because the canonical momentum $p = mv + qA$ should be used in the Sommerfeld quantum condition (1). The velocity 
$$\oint dl v = \frac{2\pi \hbar }{m}(n - \frac{\Phi}{\Phi _{0}} ) \eqno{(3)}$$
of a particle with a charge $q$ cannot be equal to zero in this case, when the magnetic flux $\Phi = \oint_{l}dl A$ inside the ring is not divisible $\Phi \neq n\Phi _{0}$ by the flux quantum $\Phi _{0} = 2\pi \hbar /q$. The persistent current of electrons \cite{PCScien09,PCPRL09} is created by one electron at the Fermi level $n _{F}$ for each one - dimensional channel, since electrons, being fermions, occupy levels from $n = -n_{F}$ to $n = +n_{F}$ with the opposite direction of velocity (3) \cite{PC1988Ch1}. Electrons at the Fermi level have a very large quantum number $n_{F} \gg 1$. Therefore, the persistent current of electrons is observed in rings with $r \approx  300 \div 800 \ nm$ at $T \approx (2n_{F}+1)\hbar ^{2}/2mr^{2}k _{B} \approx 1 \ K$ \cite{PCScien09,PCPRL09} rather than at $T \approx \hbar ^{2}/2mr^{2}k _{B} \approx 0.001 \ K$.  

The persistent current observed in \cite{PCScien09,PCPRL09} oscillates in the magnetic field $B$ with a period $B_{0} = \Phi _{0}/S$ corresponding to the flux quantum $\Phi _{0} = 2\pi \hbar /q$ inside the ring with the area $S = \pi r^{2}$. The magnitude of the flux quantum $\Phi _{0}  = 2\pi \hbar /q \approx 41.4  \ Oe \ \mu m^{2}$ corresponds the charge of electron $q = e$. These quantum oscillations are observed at very low temperature corresponding to the intermediate discreteness $\Delta E \approx k_{B}T$ in accordance with the correspondence principle. Their amplitude decreases exponentially with increasing of the temperature $T$ or the ring radius $r$ \cite{PCScien09}. The amplitude decreases exponentially from  $I_{p,A} \approx  1 \ nA = 10^{-9} \ A$ at $T \approx  0.3 \ K$ to $I_{p,A} \approx  0.01 \ nA = 10^{-11} \ A$ at $T \approx  2.5 \ K$ when the ring radius $r \approx  300 \ nm$ and  from  $I_{p,A} \approx  0.005 \ nA = 5 \ 10^{-12} \ A$ at $T \approx  0.3 \ K$ to $I_{p,A} \approx  0.0008 \ nA = 8 \ 10^{-13} \ A$ at $T \approx 0.4 \ K$ when $r \approx  800 \ nm $ \cite{PCScien09}. 

\section{The persistent currents of Cooper pairs contradict the correspondence principle}
 \label{}    
The persistent current observed in superconducting rings \cite{PCJETP07,JETP07J}, unlike the one observed in normal metal rings \cite{PCScien09,PCPRL09}, corresponds to strong discreteness $\Delta E \gg k_{B}T$ \cite{NanoLet2017} rather than intermediate discreteness $\Delta E \approx k_{B}T$. In both cases, the persistent current oscillates in a magnetic field. But the period of the oscillations $\Phi _{0} = 2\pi \hbar /q \approx 20.7 \ Oe \ \mu m^{2}$ in a superconducting ring corresponds to the charge of a pair of electrons $q = 2e$ rather than one electron $q = e$. This experimental result indicates that the mobile charge carriers in a superconductor are Cooper pairs \cite{BCS1957}. 

Strong discreteness occurs in atoms due to the small radius of electron orbits $r_{B} \approx 0.05 \ nm = 5 \ 10^{-11} \ m$. But how can be observed strong discreteness in rings with a large radius $r \approx  2 \ \mu m = 2 \ 10^{-6} \ m$ \cite{PCJETP07,JETP07J}? Neither Bohr's quantization nor the Schrodinger wave mechanics can explain this experimental fact. But the Ginzburg-Landau (GL) theory \cite{GL1950} successfully describes the macroscopic quantum phenomena of the persistent current observed in superconducting rings \cite{PCJETP07,JETP07J}. The GL expression for the superconducting current density
$$j = \frac{q}{m}n_{s}(\hbar \nabla \varphi - qA) \eqno{(4)}$$
of particles with a charge $q$, a mass $m$ and a density $n_{s}$ makes it possible to describe macroscopic quantum phenomena observed in superconductors. The relationship  
$$\mu _{0}\oint_{l}dl \lambda _{L}^{2} j  + \Phi = n\Phi_{0}  \eqno{(5)}$$  
between the current density $j$ along any closed path $l$, the magnetic flux $\oint_{l}dl A = \Phi $ inside this path and the quantum number $n$ can be deduced from (4) using the requirement $\oint_{l}dl \nabla \varphi = n2\pi $ that the GL wave function $\Psi _{GL} =|\Psi _{GL}|\exp{i\varphi }$ must be single-valued at any point in the superconductor. Here $\lambda _{L} = (m/\mu _{0}q^{2}n_{s})^{0.5}$ is a value commonly referred to as the London penetration depth. 

The persistent current in a superconducting ring with a small cross section $s \ll \lambda _{L}^{2}$ used in \cite{PCJETP07,JETP07J} must be equal to
$$I_{p} = sj = \frac{\Phi_{0}}{L_{k}}(n - \frac{\Phi }{\Phi_{0}}) = sqn_{S}\frac{\hbar }{mr}(n - \frac{\Phi }{\Phi_{0}}) \eqno{(6)}$$ 
since the current density $j$ does not change along $s$ in this case. $L_{k} = (\lambda _{L}^{2}/s)\mu_{0}2\pi r = m2\pi r/sq^{2}n_{s} $ is the kinetic inductance of a ring with radius $r$, cross-section $s$ and density $n_{s}$ of superconducting particles with charge $q$. The spectrum of  permitted values of the kinetic energy of the persistent current
$$E_{k} = \frac{L_{k}I_{p}^{2}}{2} =  I_{p,A}\Phi_{0}(n - \frac{\Phi }{\Phi_{0}})^{2} \eqno{(7)}$$ 
is strongly discrete at the values of the amplitude $I_{p,A} = \Phi_{0}/2L_{k}$ of the oscillation of the persistent current experimentally observed $I_{p,A}(T) \approx  200 \ \mu A (1 - T/T_{c})$ in rings with a radius $r \approx 2 \ \mu m$ and a cross-section $s \approx  0.01 \div 0.02 \ \mu m^{2} $ \cite{PCJETP07}. The energy difference $|E_{k}(n+1) - E_{k}(n)| \approx 0.6 I_{p,A}\Phi_{0}$ between the permitted states $n+1$ and $n$ at the magnetic flux $\Phi \approx (n+0.2)\Phi_{0}$ inside the superconducting ring corresponds to the temperature $T \approx 0.6 I_{p,A}\Phi_{0}/k_{B} \approx 1000 \ K$ at $I_{p,A} \approx 10 \ \mu A$ and $T \approx 10000 \ K$ at $I_{p,A} \approx 100 \ \mu A$. Thus, the discreteness of the spectrum in superconducting rings with the radius $r \approx 2 \ \mu m = 2000  \ nm $ is almost the same as in an atom with the radius $r_{B} \approx 0.05 \ nm $, according to the GL theory \cite{GL1950} and experimental results \cite{PCJETP07}, and contrary to the correspondence principle.

The strong discreteness $\Delta E \gg k_{B}T$ of the energy spectrum (7) is experimentally corroborated by the observed predominate probability $P(n) \propto \exp -E_{k}(n)/k_{B}T$ of a superconducting state with minimal kinetic energy (7). The measured parameters \cite{PCJETP07,JETP07J,NanoLet2017,Letter2003,Physica2019,PLA2012Ex,APL2016,PLA2017,PLA2020,Letter2007} associated with the persistent current $I_{p}$ (6) periodically change in a magnetic field with a period corresponding to the flux quantum $\Phi _{0}$ inside the ring area $S = \pi r^{2}$ due to a change in the quantum number $n$ corresponding to the minimal kinetic energy (7). Measurements of the critical current \cite{PCJETP07,JETP07J} indicate that the quantum number $n$ takes the same integer number $n$ at a given value of the magnetic flux $\Phi = BS$ at each transition to the superconducting state \cite{PRB2014}. The observation of two states $n$ and $n+1$ in rare cases, see Fig.1 in \cite{PLA2020}, only emphasizes the strong discreteness $\Delta E \gg k_{B}T$ of the permitted state spectrum of superconducting rings.  

The strong discreteness $\Delta E \gg k_{B}T$ is also confirmed by the absence of an exponential decrease of the persistent current with increasing temperature. The amplitude $I_{p,A} = \Phi_{0}/2L_{k} = (q\hbar s/mr)n_{s} $ of the persistent current of Cooper pairs depends on temperature $I_{p,A}(T) = I_{p,A}(0)(1 - T/T_{c})$ \cite{PCJETP07} only due to the temperature dependence of the density of Cooper pairs $n_{s}(T) = n_{s}(0)(1 - T/T_{c})$ in accordance with the conventional theory of superconductivity \cite{BCS1957,GL1950}. The persistent current of Cooper pairs with the amplitude $I_{p,A} \approx  0.1 \ \mu A = 10^{-7} \ A$ is observed even in the fluctuation region near the superconducting transition $T \approx  T_{c}$, where the resistance is not zero $R > 0$ \cite{Letter2007,Science2007}. The first experimental evidence of this paradoxical phenomenon was obtained in 1962 by W.A. Little and R.D. Parks \cite{LP1962}. The Little - Parks oscillations are observed both in aluminum rings at $T \approx  T_{c} \approx  1.35 \ K$ \cite{Letter2007} and high-temperature superconductor (HTSC) rings \cite{LP2010Nature,LP2010PRB} at temperature up to $T \approx  T_{c} \approx  86 \ K$ \cite{LP1990PRB}.

Expression (5) of the GL theory \cite{GL1950} can also describe the quantization of the magnetic flux $\Phi = n\Phi_{0}$, which was first observed in superconducting cylinders with thick walls $w \gg \lambda _{L}$ in 1961 \cite{fluxquan1961a,fluxquan1961b}. The integral in (5) along the closed path $l$ inside the wall should be zero $\oint_{l}dl j = 0$ in this case of the strong screening and, therefore, $\Phi = n\Phi_{0}$ in accordance with (5). The persistent current equal $I_{p} \leq j_{c}\lambda _{L}L \approx 40 \ A$ for the length $L \approx 8 \ mm = 8 \ 10^{-3} \ m  \gg \lambda _{L}$ of the cylinder used in \cite{fluxquan1961a}, the typical values of the London penetration depth $\lambda _{L} \approx 50 \ nm = 5 \ 10^{-8} \ m$ and of the critical current density $j_{c} \approx 10^{11} \ A/m^{2}$ in a superconducting cylinder with the strong screening $w \gg \lambda _{L}$ is significantly greater, than the persistent current $I_{p} \leq j_{c}s$ in a ring with the weak screening $s \ll \lambda _{L}^{2}$. The persistent current $I_{p} \approx 40 \ A$ corresponds to the macroscopic angular momentum $P_{r} = I_{p}2\pi r^{2}(m/e) \approx  10^{15} \ \hbar $ for the radius $r \approx 10 \ \mu m $ of the cylinder used in \cite{fluxquan1961a}.  

The GL theory \cite{GL1950} describes the Meissner effect \cite{Meissner1933} as a special case of the flux quantization $\Phi = n\Phi_{0} $ when $n = 0$. The quantum number $n$ can be non-zero in (5) if only a singularity of the GL wave function $\Psi _{GL} =|\Psi _{GL}|\exp{i\varphi }$ is inside the closed path $l$. A hole in the superconducting cylinder \cite{fluxquan1961a,fluxquan1961b} and the Abrikosov vortex \cite{Abrikosov} are such singularities. The integral $\oint_{l}dl \nabla \varphi = n2\pi = 0$ must be zero if the closed path $l$ can be reduced to a point in the region inside $l$ without singularity. Thus, according to the GL theory \cite{GL1950} quantization forces mobile charge carriers in a superconductor to move in the direction opposite to the electromagnetic force at the appearance of the persistent current in the cases of a ring with the weak screening $s \ll \lambda _{L}^{2}$ (6) \cite{PCJETP07,JETP07J}, the flux quantization \cite{fluxquan1961a,fluxquan1961b} and of the Meissner effect \cite{Meissner1933}. The persistent current must appear when superconducting state with zero current density is forbidden in accordance with the expression (5) of the GL theory \cite{GL1950}. 

According to the GL theory \cite{GL1950}, the superconducting transition is reversible in the sense that the superconductor returns to the same superconducting state in which it was before its transition to the normal state. For example, a bulk cylinder with a macroscopic radius $R \gg \lambda_{L}$ returns to the superconducting state with the same density of the surface screening current $j = (H/\lambda_{L})\exp (r - R)/\lambda_{L}$ and with the same magnetic flux density $B = \mu _{0}H\exp (r - R)/\lambda_{L}$ when cooled in the same magnetic field $H$. The Meissner effect \cite{Meissner1933} is the experimental evidence of this reversibility. The magnetic flux quantization \cite{fluxquan1961a,fluxquan1961b} is the  experimental evidence of the reversibility in the case of cooling in an external magnetic field of a hollow cylinder \cite{fluxquan1961a} or ring \cite{fluxquan1961b} with the strong screening $w \gg \lambda_{L}$. The results of the measurements of the critical current \cite{PCJETP07,JETP07J} are one of the experimental evidences of the reversibility in the case of rings with weak screening $s \ll \lambda _{L}^{2}$. The ring is switched between the superconducting and normal states by the external measuring current at each measurement of the critical current \cite{PRB2014}. The results of all measurements testify that the ring returns to the superconducting state with the same quantum number and the same persistent current at a given value of the external magnetic field \cite{PCJETP07,JETP07J}. 

\section{The puzzle with the conservation law is a consequence of a violation of the correspondence principle}
\label{}  
Hirsch believes that the contradiction with the conservation law is the puzzle of the theory of superconductivity. But a comparison of the persistent current of electrons and Cooper pairs reveals that this contradiction is a problem of quantum theory and a violation of the correspondence principle rather than the theory of superconductivity. The electric current $I = sen_{e}\overline{v}$ decays during a short relaxation time $\tau_{RL} = L /R$ in a normal metal ring with the resistance $R = 2\pi r\rho /s$ and the inductance $L$ without an electric field $2\pi rE = -d\Phi /dt = 0$. The resistance is not zero in a normal metal because of electron scattering which, reduces the average electron velocity $\overline{v}$ to zero in the absence of an electromotive force $F_{e} = eE = 0$. The average velocity can be non-zero $\overline{v} \neq 0$ during a long time $t \gg \tau_{RL}$ due to the force balance $F_{e} + F_{dis} = 0$, when the dissipation force $F_{dis} \neq 0$, which is non-zero due to electron scattering, is balanced by the electromotive force $F_{e} = eE \neq 0$. 

But in the rings with non-zero resistivity $\rho \approx 10^{-8} \ \Omega m $ and with an elastic electron scattering length $l_{sc} \approx 40 \ nm$ that is much smaller than the circumference of the ring $2\pi r \geq  1800 \ nm$ \cite{PCScien09}, the persistent current \cite{PCScien09,PCPRL09} does not decay for a long time $t \gg \tau_{RL}$ at $F_{e} = eE = 0$ . This paradoxical phenomenon is observed due to the Sommerfeld quantization condition (1). The average velocity should be equal to zero: $\overline{v} = 0$ after scattering of electron, when its coordinate becomes more certain, the momentum becomes less certain and the average value of the angular momentum should be equal $\overline{p}r = \oint_{l}dl \overline{p}/2\pi = \oint_{l}dl (m\overline{v} + eA)/2\pi = e\Phi /2\pi = \hbar \Phi /\Phi_{0} $. But electrons can return from time to time to the state (1) with a certain angular momentum $pr = \oint_{l}dl p/2\pi  = n \hbar $ and uncertain coordinate. 

Thus, the angular momentum of the electron should change from $pr = n \hbar $ to $\overline{p}r = \hbar \Phi /\Phi_{0} $ due to scattering and from $\overline{p}r = \hbar \Phi /\Phi_{0} $ to $pr = n \hbar $ due to quantization (1). The change in momentum $\Delta \overline{p} = m\Delta \overline{v} = \hbar(n - \Phi /\Phi_{0})/r$ due to the quantization per unit of time $\Delta \overline{p} f_{re} = \hbar(n - \Phi /\Phi_{0})f_{re}/r = F_{q}$ was called 'quantum force' $F_{q}$ in the article \cite{PRB2001}. Here $f_{re}$ is the frequency of the return of the electron to the quantum state (1) with a certain angular momentum. The change of the momentum $\Delta \overline{p} = \hbar(\Phi /\Phi_{0} - n)/r$ because of the electron scattering in a unit time $\Delta \overline{p} f_{re} = \hbar(\Phi /\Phi_{0} - n)f_{re}/r = F_{dis}$ is the dissipation force $F_{dis}$. We can write the force balance $F_{q} + F_{dis} = 0$ explaining why the persistent current does not decay \cite{PCScien09,PCPRL09}, despite the non-zero resistance $R > 0$ and the zero electromotive force $F_{e} = eE = 0$.

It should be emphasized that the quantum force used in the article \cite{PRB2001} does not explain, but only describes the change in angular momentum  due to quantization (1) per unit time and cannot claim to explain the Meissner effect, contrary to Hirsch's assumption \cite{Hirsch10Meis}. The change in the angular momentum of electrons due to the quantization (1) could cause a contradiction between quantum theory and the conservation law if quantum mechanics claimed on an exact space-time description of quantum phenomena. Bohr wrote about this in an article \cite{Bohr1936}: {\it "Any attempt at an accurate space-time description of quantum phenomena implies a rejection of the strict application of conservation laws . . . Conversely, the strict application of the conservation laws to quantum phenomena implies a significant limitation of the accuracy of the space-time description}. The same statement is in Bohr's 1958 publication: {\it "Any strict application of the laws of momentum and energy conservation to atomic processes presupposes the rejection of certain localization of particle in space and time. This statement is represented quantitatively with the Heisenberg's uncertainty relation"} \cite{Bohr1958}.

Quantum mechanics does not claim to be an accurate space-time description of quantum phenomena. Therefore, the persistent current of electrons, which was first reliably observed in 2009 \cite{PCScien09,PCPRL09}, does not contradict the conservation law according to quantum mechanics, despite the fact that it does not decay at nonzero resistance $R > 0$ and in the absence of an electromotive force $F_{e} = eE = 0$. The persistent current of Cooper pairs, undamped at nonzero resistance $R > 0$, was first observed much earlier, in the 1962 Little-Parks experiment \cite{LP1962}. Hirsch rightly wrote \cite{Hirsch10Meis} that the 'quantum force' was used in 2001 \cite{PRB2001} in order to explain the Little-Parks effect \cite{LP1962}, which was observed in numerous experiments \cite{Letter2007,LP2010Nature,LP2010PRB,LP1990PRB}. The persistent current of Cooper pairs $I_{p} \neq 0$ is observed at nonzero resistance $R > 0$ \cite{Letter2007,Science2007,LP1962,LP2010Nature,LP2010PRB,LP1990PRB} only in a narrow temperature range near the superconducting transition $T \approx T_{c}$. In this range, the electric current $\overline{I_{p}} \neq 0$ does not decay at $\overline{R} > 0$, since thermal fluctuations \cite{Skocpol1975} switch the ring or its segments between the normal state with $n_{s} = 0$ and superconducting state with $n_{s} > 0$: the current decays at $n_{s} = 0$ and $R > 0$ due to the dissipation force $F_{dis}$ and returns to a non-zero value $I_{p} = sqn_{s}v$ at $n_{s} > 0$ and $R = 0$ due to the quantization (3) \cite{PRB2001}.   

This explanation \cite{PRB2001} does not go beyond the conventional theory of superconductivity \cite{BCS1957,GL1950}, according to which the persistent current (6) $I_{p} = sqn_{s}v$ emerges in the superconducting state due to the quantization (3) and is damped in the normal state due to a non-zero resistance $R > 0$ with the generation of Joule heat. The observation of the persistent current $\overline{I_{p}} \neq 0$ in rings with a non-zero resistance both of electrons \cite{PCScien09,PCPRL09} and of Cooper pairs \cite{Letter2007,Science2007,LP1962,LP2010Nature,LP2010PRB,LP1990PRB} can be explained by the force balance $F_{q} + F_{dis} = 0$ \cite{PLA2012T,PhysicaC2021}. But there is an important difference. The change in the angular momentum of electrons due to the quantization (1) is based on the Heisenberg uncertainty relation and cannot exceed  Planck's constant $\hbar (n - \Phi /\Phi_{0}) \leq \hbar$. Such microscopic changes $\Delta p_{r} \leq \hbar$ are observed in many quantum phenomena, for example at measurement of spin projection \cite{ConserLaw2010}. 

According to GL theory \cite{GL1950}, all Cooper pairs in a ring, unlike to electrons, have the same quantum number $n$ describing their angular momentum (1). For this reason, the angular momentum of mobile charge carriers changes not by a microscopic $\hbar (n - \Phi /\Phi_{0}) \leq \hbar$, but by a macroscopic amount $N_{s}\hbar (n - \Phi /\Phi_{0}) \gg  \hbar$, since the number of Cooper pairs $N_{s} = 2\pi r s n_{s}$ in a real superconducting ring is very large. For example, the persistent current $I_{p} \approx 10 \ \mu A$ observed in a ring with a radius $r \approx 2 \ \mu m$ and a cross section $s \approx  0.01 \ \mu m^{2} $ \cite{PCJETP07} is created by $N_{s} \approx 5 \ 10^{7}$ Cooper pairs. The appearance of this persistent current is as much a puzzle as the Meissner effect since it corresponds to a macroscopic change in the angular momentum $P_{r} = I_{p}2\pi r^{2}(m/e) \approx  1.5 \ 10^{-27} J s \approx  10^{7} \ \hbar $ without any force. But in this case, it is more obvious that the reason for the contradiction with Faraday's law and the law of angular momentum conservation is quantization (3): the angular momentum changes by the macroscopic amount $\Delta P_{r} = N_{s}\hbar (n - \Phi /\Phi_{0}) \approx 10^{7} \ \hbar \gg \hbar$, since superconducting state with zero velocity of Cooper pairs (3) is forbidden due to the quantization (1). 

The observations \cite{NanoLet2017,nJump2002Geim,nJump2007Moler,nJump2016Nature} of macroscopic jumps of the persistent current confirm that all Cooper pairs in a ring have the same quantum number $n$. The persistent current increases under the influence of the Faraday electrical field $L_{k}I_{p} = E 2\pi r = -d\Phi /dt = -d(BS - LI_{p})/dt $ with the change of the external magnetic field $B$ and decreases by the macroscopic amount $\Delta I_{p} = \Delta n \Phi_{0}/L_{k}$ at a critical value of the superconducting current $I_{p} = I_{c}$ \cite{NanoLet2017,nJump2002Geim,nJump2007Moler,nJump2016Nature} due to a change of the quantum number $\Delta n $ by one $\Delta n = 1$ \cite{NanoLet2017,nJump2016Nature} or by other integers $\Delta n = 2,3,4...$ \cite{nJump2002Geim,nJump2007Moler}. The jump of the persistent current $\Delta I_{p} = 150 \ \mu A$ observed in \cite{nJump2016Nature} corresponds to the change of the angular momentum on the macroscopic value $\Delta P_{r} \approx 2 \ 10^{8} \ \hbar $. 

Thus, the microscopic quantum theory, that is, the Bohr quantization and the Schrodinger wave mechanics, does not contradict the conservation law due to the correspondence principle, according to which the change in angular momentum due to the quantization (1) cannot exceed Planck's constant $\Delta p_{r} \leq \hbar$. The creators of the conventional theory of superconductivity \cite{BCS1957,GL1950} had to violate the correspondence principle, since according to this principle no macroscopic quantum phenomena can be observed. This macroscopic quantum theory explains the macroscopic change in the angular momentum $\Delta P_{r} \gg \hbar $ of mobile charge carriers observed in the Meissner effect \cite{Meissner1933} and other macroscopic quantum phenomena \cite{LP1962,fluxquan1961a,fluxquan1961b} as a consequence of the quantization (1): the angular momentum must change on a macroscopic amount during the transition to a superconducting state when states with zero superconducting current (4) are forbidden in accordance with the GL expression (5).   
    
\section{The Schrodinger wave function and the Ginzburg-Landau wave function}
 \label{}
From the very beginning, it was assumed that the Meissner effect is a macroscopic quantum phenomenon since it contradicts the laws of classical physics. Lev Landau was the first to propose in 1941 the theory of the Meissner effect as a quantum phenomenon in an article \cite{Landau41} devoted mainly to the theory of superfluidity of $^{4}He$ liquid, that is helium II. Landau proposed to describe the Meissner effect using of a wave function \cite{Landau41}, which was later used in the GL theory \cite{GL1950}. He deduced from this wave function an expression for the density of superconducting current (4), which allows not only to describe the Meissner effect, discovered in 1933, but also to predict the magnetic flux quantization, discovered in 1961 \cite{fluxquan1961a,fluxquan1961b}, and quantum oscillation of the persistent current, discovered in 1962 \cite{LP1962}. L.D. Landau followed to the Schrodinger wave mechanics \cite{Schrodinger1926}. But his theory \cite{Landau41} would not be able to macroscopic quantum phenomena if his wave function did not differ from Schrodinger's one, since Schrodinger's wave mechanics satisfies the correspondence principle.   

In fact, L.D. Landau explained in the beginning of his 1941 paper why microscopic quantum theory, such as Schrodinger's wave mechanics, cannot describe macroscopic quantum phenomena: "{\it Tisza \cite{Tisza1938} proposed to consider helium II as a degenerate ideal Bose gas. It is assumed that atoms in the ground state (a state with zero energy) move in a liquid without friction either against the walls of the box or against the rest of the liquid. However, such a representation cannot be considered satisfactory. Not to mention the fact that liquid helium has nothing to do with an ideal gas, atoms in the ground state would not behave like "superfluids" at all. On the contrary, nothing could prevent atoms in a normal state from colliding with excited atoms, i.e., when moving through a liquid, they would experience friction and there would be no question of any "superfluidity". Thus, the explanation proposed by Tissa is only apparent, and not only does not follow from the assumptions made, but directly contradicts them}" \cite{Landau41}.   

Schrodinger's wave mechanics considers a degenerate ideal Bose gas in a box with macroscopic sizes $L_{x}$, $L_{y}$, $L_{z}$ as the states of atoms with a mass $m$. The kinetic energy of the atom $E_{k} = p_{x}^{2}/2m + p_{y}^{2}/2m + p_{z}^{2}/2m = n_{x}^{2}h^{2}/2mL_{x}^{2} + n_{y}^{2}h^{2}/2mL_{y}^{2} + n_{z}^{2}h^{2}/2mL_{z}^{2}$ is determined by its mass $m$, by the macroscopic sizes $L_{x}$, $L_{y}$, $L_{z}$ of the box and by the integer numbers $n_{x}$, $n_{y}$, $n_{z}$. Even if all the atoms of the Bose-Einstein condensate are at the lowest level $n_{x} = 0$, $n_{y} = 0$, $n_{z} = 0$, they require negligible energy $h^{2}/2mL_{x}^{2}$, or $h^{2}/2mL_{y}^{2}$, or $h^{2}/2mL_{z}^{2}$ to move to the next level $n_{x} = 1$, or $n_{y} = 1$, or $n_{z} = 1$. Consequently, L.D. Landau \cite{Landau41} was right while L. Tisza \cite{Tisza1938} and also F. London \cite{London1938}, who proposed the Bose-Einstein condensate as a mechanism for superfluidity in $^{4}He$ and superconductivity, were not right. 

The energy difference between adjacent permitted states $n_{x}$, $n_{y}$, $n_{z}$ in a box with macroscopic sizes $L_{x}$, $L_{y}$, $L_{z}$ should be negligible and  correspond to a continuous spectrum $h^{2}/2mL_{x}^{2} \ll k_{B}T$, $h^{2}/2mL_{y}^{2} \ll k_{B}T$ $h^{2}/2mL_{z}^{2} \ll k_{B}T$ when microscopic particles can move separately, as microscopic quantum mechanics assumes. Landau considered the motion of helium II liquid, rather than of individual helium II atoms, to create his theory of superfluidity of $^{4}He$ liquid \cite{Landau41}. He actually had to postulate that helium II atoms cannot move separately in order to explain the superfluidity as a consequence of an energy gap in the excitation spectrum of helium II. 

Landau used this postulate at the end of the article \cite{Landau41} to describe the Meissner effect. In 1941, he derived the GL wave function and the GL expression (4) based on this postulate. But electrons, unlike $^{4}He$ atoms, are fermions, not bosons and, according to the fundamentals of quantum mechanics, cannot be on the same level. In the late forties and early fifties of the last century, many theorists tried to explain how fermions can turn into bosons. Almost all experts, except Hirsch, believe that the efforts of theorists were crowned with success in 1957, when J. Bardin, L.N. Cooper and J. R. Schrieffer showed that electrons can form Cooper pairs, which are bosons with zero spin \cite{BCS1957}. The BCS theory \cite{BCS1957} confirmed Landau's assumption \cite{Landau41} that mobile charge carriers in a superconductor can be on the same level.  The GL wave function differs from the Schrodinger wave function due to Landau's postulate. Schrodinger's wave mechanics consider the individual movement of microparticles. The square of the amplitude $|\Psi |^{2}$ of the Schrodinger wave function is interpreted as the probability of finding an individual microparticle, such as an electron, in unit volume \cite{FeynmanL}. Unlike the Schrodinger wave mechanics, the square of the amplitude $|\Psi _{GL}|^{2} = n_{s}$ is the density of all Cooper pairs according to the GL theory \cite{GL1950}.

Richard Feynman drew attention to this fundamental difference between microscopic quantum mechanics and the theory of superconductivity in the Section "The Schrodinger Equation in a Classical Context: A Seminar on Superconductivity" of his Lectures on Physics \cite{FeynmanL}. He stated that Schrodinger "{\it imagined incorrectly that $|\Psi |^{2}$ was the electric charge density of the electron. It was Born who correctly (as far as we know) interpreted the $\Psi $ of the Schrodinger equation in terms of a probability amplitude - that very difficult idea that the square of the amplitude is not the charge density but is only the probability per unit volume of finding an electron there, and that when you do find the electron some place the entire charge is there}". But Feynman further wrote that "{\it in a situation in which $\Psi $ is the wave function for each of an enormous number of particles which are all in the same state, $|\Psi |^{2}$ can be interpreted as the density of particles}" \cite{FeynmanL}. 

This fundamental difference is especially evident when comparing the theory of the persistent current of electrons \cite{PC1988Ch1} and Cooper pairs \cite{GL1950}. One electron at the Fermi level creates a current $I _{p,1} \approx ev_{F}/2\pi r $ in a one - dimensional channel \cite{PC1988Ch1} in accordance with the Schrodinger wave function, where $e/2\pi r = |\Psi |^{2}$ can be interpreted both as the electric charge density of electrons and as the probability per unit volume. The total persistent current $I _{p,t} $ in the normal metal ring is created by the currents $I _{p,1} $ of a large number $N _{ch}$ of one - dimensional channels. The $I _{p,1} $ direction in these channels is random. Therefore, the persistent current of electrons in the ring $I _{p,t} \approx \surd N _{ch}I _{p,1}$ is proportional to $\surd N _{ch}$ rather than $N _{ch}$. The persistent current in a superconducting ring \cite{PCJETP07} is much greater than in the normal metal ring \cite{PCScien09,PCPRL09} since it should be equal $I _{p} = sq|\Psi _{GL}|^{2}v = qN_{s}v/2\pi r$, according to the GL theory \cite{GL1950}, where $N_{s}  = \int_{V}dV |\Psi |^{2} = \oint_{l}dl sn_{s} = 2\pi r sn_{s}$ is the total number of Cooper pairs in the ring with the section $s$ and the circumference $l = 2\pi r$; $v = (\hbar \nabla \varphi - qA)/m$ is the velocity of $N_{s}$ Cooper pairs. 

However, the GL theory \cite{GL1950} would not be able to describe superconductivity as a macroscopic quantum phenomenon if its function describes only superconducting condensate with macroscopic mass $M = mN_{s}$. In this case, the spectrum of a superconducting ring with a macroscopic volume $V = s2\pi r$ and a macroscopic number of Cooper pairs could not be discrete in accordance with Bohr's quantization (2). The energy spectrum of a superconducting ring is strongly discrete (7), since in accordance with the GL theory, on one hand, the velocity $v = p/m $ is determined by the microscopic mass $m$ of each Cooper pair with a microscopic mass $m$ and charge $q$, but, on the other hand, the macroscopic mass $M = mN_{s} = mV|\Psi _{GL}|^{2} = m2\pi r sn_{s}$ of superconducting condensate with a density $|\Psi _{GL}|^{2} = n_{s}$ in the entire ring volume $V = 2\pi r s$ is used in the expression for the kinetic energy $Mv^{2}/2$. As a consequence, the energy difference $\Delta E = E_{n+1} - E_{n}$ between adjacent permitted states $n$ and $n+1$ of a superconducting ring  
$$\Delta E = \frac{Mv_{n+1}^{2}}{2} - \frac{Mv_{n}^{2}}{2} = N_{s}(2n+1)\frac{\hbar ^{2}}{2mr^{2}} \eqno{(8)}$$
is more by a multiplier equal to the total number of Cooper pairs $N_{s}$. 

This number is very large $N_{s}  = 2\pi r sn_{s} \gg 1$ in a real ring with a cross section $s \ll \lambda _{L}^{2}$ and a circumference $l = 2\pi r$. The discreteness (8) increases with an increase in all three sizes of the ring $\Delta E \approx N _{s}\hbar ^{2}/2mr^{2} \approx  n_{s}s2\pi r(\hbar ^{2}/2mr^{2}) \propto  (s/r)$ due to an increase in the number of Cooper pairs $N _{s}$. Only the kinetic energy is taken into account, "{\it Since the total energy due to the field term $h^{2}/8\pi$ is less than the kinetic energy of the current by a factor of the order of the ratio of the cross-sectional area of the conductor to $\lambda _{L}^{2}$}" \cite{Tinkham}. The kinetic energy $E_{k} = L_{k}I_{p}^{2}/2$ (7) exceeds the magnetic field energy $E_{f} = L_{f}I_{p}^{2}/2$ generated by the persistent current $I_{p}$ at $s \ll \lambda _{L}^{2}$, since the kinetic inductance $L _{k} \approx (\lambda _{L}^{2}/s) \mu _{0}2\pi r$ exceeds the magnetic inductance $L_{f}  \approx \mu _{0}2\pi r$ in this case of weak screening $E_{k}/E_{f} = L_{k}/L_{f} \approx \lambda _{L}^{2}/s \gg 1$. 

\section{Contradiction with the second law of thermodynamics}
\label{}
We cannot doubt that the persistent current (6) appears in the ring during its transition to the superconducting state due to quantization (1), since this is not only a result of the GL theory \cite{GL1950}, but also an experimental fact. But how can the persistent current disappear when the ring reverts to the normal state? Superconductivity experts, including the creators of the conventional theory of superconductivity \cite{BCS1957,GL1950}, did not consider this as a problem for many years, since it was obvious that in the normal state the electric current is damped due to non-zero resistance, and its kinetic energy is dissipated into Joule heat. They stopped understanding what physicists understood ninety years ago. The famous physicist W.H. Keesom wrote in 1934, after discovery of the Meissner effect, that "{\it it is essential that the persistent currents have been annihilated before the material gets resistance, so that no Joule heat is developed}" \cite{Keesom1934}. 

Before the discovery of the Meissner effect in 1933, W.H. Keesom and other physicists thought that the transition of a bulk superconductor to the normal state in a magnetic field is irreversible. A well-known expert on superconductivity, D. Shoenberg, wrote in 1952: "{\it At that time it was assumed that the transition in a magnetic field is substantially irreversible, since the superconductor was considered to be a perfect conductor (in the sense discussed in Chapter II), in which, when the superconductivity is destroyed, the surface currents associated with the field are damped with the generation of Joule heat}" \cite{Shoenberg1952}. The generation of Joule heat is an irreversible process according to the second law of thermodynamics. Therefore, if Joule heat is generated at the transition to the normal state then the Meissner effect contradicts to the second law of thermodynamics. 

W.H. Keesom actually set the task \cite{Keesom1934} for the future theory of superconductivity to explain how the persistent currents can be annihilated before the material gets resistance in order that this theory did not contradict the second law of thermodynamics. The creators of the conventional theory of superconductivity \cite{BCS1957,GL1950} did not even consider the Keesom task \cite{Keesom1934}. They, like most other superconductivity experts, kind of forgot that Joule heating is an irreversible thermodynamic process. For this reason, for many years no one noticed that the conventional theory of superconductivity \cite{BCS1957,GL1950} contradicts the second law of thermodynamics \cite{Entropy2022}, until in 2020 Hirsch recalled \cite{HirschEPL,HirschIJMP,HirschPhys} that the generation of Joule heat is an irreversible thermodynamic process. 

This contradiction, as and the contradiction with the law of angular momentum conservation and Faraday's law, is a consequence of a violation of the correspondence principle: the macroscopic persistent current (6) emerges because of the quantization (3) when the number of Cooper pairs in the ring increases from $N_{s}  = 0$ to $N_{s}  = 2\pi r sn_{s} \gg 1$ during its transition to a superconducting state. According to the conventional opinion of most experts, when the density of Cooper pairs decreases from $N_{s}  = 2\pi r sn_{s} \gg 1$ to $N_{s}  = 0$, upon returning to the normal state, their mechanical momentum is transferred to the normal electrons and "{\it this momentum then decays with the transport relaxation time}" \cite{Eilenberger1970} due to the usual scattering processes. The latter process is irreversible according to the second law of thermodynamics, and therefore the first process contradict this law \cite{Entropy2022}. 

For many years, no one noticed that the conventional theory of superconductivity \cite{BCS1957,GL1950} contradicts the laws of thermodynamic, since in our understanding of physics there is not only progress, but also regression. Most modern physicists believe that the second law of thermodynamics cannot be violated, but most of them do not know what their confidence is based on. Physicists used to know this much better. Einstein understood that thermodynamics is nothing more than a systematic answer to the question: what should be the laws of nature in order for a perpetual motion machine to be impossible. The centuries-old faith in the impossibility of a perpetual motion machine has played an important role in the history of physics. Even Einstein did not question this faith. Therefore, he was sure that "{\it Classical thermodynamics is the only physical theory of universal content, in respect which I am convinced that, within the framework of applicability of its basic concepts, it will never by overthrown}" \cite{Einstein1949}. 

Although no centuries-old faith can have a scientific basis, if only because our ideas about the laws of physics have changed fundamentally several times over the past centuries. M. Smoluchowski wrote in 1914: "{\it We call the Carnot principle, intuitively realized by him, the second law of thermodynamics since Clausius's time}" \cite{Smoluchowski}. Sadi Carnot based his principle on the centuries-old faith in the impossibility of a perpetual motion machine. In his brilliant work of 1824, he wrote that "{\it it would be not only a perpetual motion, but also an unlimited creation of motive power without the cost of phlogiston or any other agents}" \cite{Carnot} if the efficiency of a heat engine could exceed the maximum value 
$$\eta_{max} = 1 - \frac{T_{co}}{T_{he}}    \eqno{(9)}$$
which he postulated. In Carnot's time, heat was considered a liquid, phlogiston. But the relationship between the maximum efficiency of any heat engine and the impossibility of a perpetual motion machine is valid in various concepts of heat. Thus, we must use fuels to create a temperature difference $T_{he} - T_{co}$ between the heater $T_{he}$ and the cooler $T_{co}$ in order to get useful energy from heat due to the faith in the impossibility of a perpetual motion machine. 

When, starting from Clausius's time, heat began to consider to be a form of energy, it became clear that the possibility of a perpetual motion machine would be inevitable in accordance with the law of energy conservation if all physical processes could be reversible. Carnot, in fact, postulated that the conversion of any energy into heat is an irreversible process. The contradiction between the irreversibility postulated by the second law of thermodynamics and the reversibility of physical laws provoked a struggle between thermodynamic-energy and atomistic-kinetic worldviews in the late 19th and early 20th centuries, see \cite{Smoluchowski}. Most scientists of that time had a negative attitude to the Maxwell-Boltzmann statistical theory because of this contradiction. Proponents of the thermodynamic-energy worldview, such as Wilhelm Ostwald, Nobel Prize Winner 1909, denied even the existence of atoms and their perpetual thermal motion.  

Ludwig Boltzmann died in 1906 as an unrecognized scientist. But already in 1914 M. Smoluchowski wrote:  "{\it Thus, today the issue is viewed in a completely different way than twenty years ago. Atomistics is recognized as the foundation of modern physics in general; the second law of thermodynamics has once and for all lost its significance as an unshakable dogma, as one of the basic principles of physics}" \cite{Smoluchowski}. But the faith in the impossibility of a perpetual motion machine turned out to be so blind that the dogma changed rather than lost its significance. Only the notion of which processes should be considered irreversible has changed.

M. Smoluchowski wrote about what processes were considered irreversible in the 19th century: "{\it In the phenomena of fluctuation experimentally observed in recent years, it seems extremely strange to a proponent of classical thermodynamics that he sees with his own eyes the reverse course of processes that are usually regarded as irreversible. Because, according to the classical theory, the second law of thermodynamics should disappear if at least one process, regarded as irreversible, admits reversibility}" \cite{Smoluchowski}. The fluctuation phenomenon is, for example, the conversion of thermal energy into kinetic energy of Brownian particles. Since Smoluchowski's time no one believes that such a transformation violates the second law of thermodynamics if the kinetic energy does not exceed the energy of thermal fluctuations $k_{B}T$.  

The kinetic energy of the surface persistent currents pushing the magnetic flux out of the bulk superconductor due to the Meissner effect is much higher than the energy of thermal fluctuations $k_{B}T$. Therefore, the Meissner effect contradicts the second law of thermodynamics, if we think "{\it that the transition in a magnetic field is substantially irreversible, since $\cdot \cdot \cdot $, when the superconductivity destroyed, the surface currents associated with the field are damped with the generation of Joule heat}" \cite{Shoenberg1952}. The faith in the second law of thermodynamic forced all the experts on superconductivity to make conclusion that the Meissner effect was able to make the irreversible transition reversible. This conclusion could not be logically right since the Meissner effect, observed during the transition from the normal to the superconducting state, could not change the reverse transition from the superconducting to the normal state. According to the laws of physics known in 1933, the transition of a bulk superconductor to the normal state in the critical magnetic field $H = H_{c}(T)$ cannot be a phase transition, at least for four reasons: 

1) Because of the work 
$$A_{sn}  = \int _{t}dt IU = \int _{t}dt LH_{c}\frac{d\Phi }{dt} = LH_{c}\Phi = V\mu_{0}H_{c}^{2} \eqno{(10)}$$ 
which is performed by an electromotive force $U = d\Phi /dt$ on the solenoid, which maintains the magnetic field $H = H_{c}$ and the solenoid current $I = LH$ when the magnetic flux in a superconducting cylinder radius $R$, length $L$ and volume $V = L\pi R^{2}$ increases from $\Phi = B\pi R^{2} = 0$ in the superconducting state to $\Phi = \mu_{0}H_{c}\pi R^{2}$ in the normal state; 

2) Because half of the work $A_{sn} = V\mu_{0}H_{c}^{2}$ creates the energy of the magnetic field $E_{m} = V\mu_{0}H_{c}^{2}/2$ in the volume of the superconductor $V$ in the normal state and increases the free energy 
$$F_{nH} = F_{sH} + E_{m} = F_{sH} + \frac{V\mu_{0}H_{c}^{2}}{2} \eqno{(11)}$$ 
during the transition; 

3) Due to the generation of Joule heat at the dissipation of the Foucault currents induced by the second half of the work $A_{sn} - E_{m} = V\mu_{0}H_{c}^{2}/2$; 

4) The magnetic flux cannot be pushed out of a perfect conductor, and therefore the superconductor cannot return to its original state in a time-constant external magnetic field $H = H_{c}$ in accordance with physics known in 1933, Faraday's law and the law of angular momentum conservation. 

$H_{c}(T)$ is the critical magnetic field of a bulk superconductor at a temperature below the critical temperature $T_{1} < T_{c}$. $F_{nH}$ and $F_{sH}$ are the free energy in the normal state and superconducting state in an external magnetic field $H = H_{c}(T)$. The Meissner effect eliminated only the fourth cause and could not logically eliminate the other causes, since the effect observed during the transition from the normal state to the superconducting state could not change the transition from the superconducting state to the normal state. Although the Meissner effect contradicts Faraday's law, it cannot completely cancel this law according to which the work (10) must be performed when the magnetic flux changes from $\Phi = 0$ to $\Phi = \mu_{0}H_{c}\pi R^{2}$. 

Despite the obvious falsity of the conclusion made after the discovery of the Meissner effect, after 1933 all physicists were convinced that the superconducting transition in a magnetic field $H = H_{c}(T)$ is the first-order phase transition. In order to eliminate the contradiction with the second law of thermodynamics, in addition to this confidence, it was necessary to solve the task set by W.H. Keesom \cite{Keesom1934} - to explain how the persistent currents can be annihilated before the material gets resistance. The conventional theory of superconductivity \cite{BCS1957,GL1950} is internally inconsistent \cite{HirschEPL,HirschIJMP,HirschPhys} and contradicts the second law of thermodynamics \cite{Entropy2022}, since its creators did not solve the Keesom task \cite{Keesom1934}, i.e. did not eliminate the third reason, according to which the transition of a bulk superconductor to the normal state cannot be considered a phase transition. 

\section{The false conclusion made in 1933 provoked obvious contradictions in books on superconductivity}
\label{}
W.H. Keesom, at least, set the task of eliminating the third reason, according to which the superconducting transition in the critical magnetic field $H = H_{c}(T)$ cannot be considered a phase transition. The first and second reasons were ignored, although W.H. Keesom wrote about the work (10) $A_{sn} = V\mu_{0}H_{c}^{2}$ in 1934: "{\it Suppose we are realizing a cycle as considered by Gorter, 1) Cooling from just above the normal transition point $T_{0} > T_{c}$ to a temperature $T_{1} < T_{c}$, with external field $H = 0$, 2) applying a magnetic field $H$, just below the threshold value at $T_{1} < T_{c}$, 3) increasing the temperature to $T_{0} > T_{c}$, the field $H$ being kept constant, and 4) switching off the magnetic field $H = 0$. In passing (at the beginning of process 3) the threshold value curve $H_{c}(T)$, the electromotive force on the solenoid that maintains the magnetic field must do an amount of work equal to twice the energy of the field that comes into existence in the metal}" \cite{Keesom1934}. This cycle is called 'Gorter cycle' in some publications \cite{Hirsch2017PRB}. 

W.H. Keesom understood that the energy of the magnetic field $E_{m} = V\mu_{0}H_{c}^{2}/2$ arises only in the normal state and is zero $VBH_{c}/2 = 0$ in the superconducting state when the magnetic flux density $B = 0$ in the volume $V$ of a superconductor. He knew that the magnetic field should increase the free energy of the normal state 
$$F_{nH}(T_{0}) = F_{n0}(T_{0}) + E_{m} = F_{n0}(T_{0}) + \frac{V\mu_{0}H^{2}}{2} \eqno{(12)}$$
and should not influence on the free energy of the superconducting state  
$$F_{sH}(T_{1}) = F_{s0}(T_{1}) \eqno{(13)}$$
W.H. Keesom also knew that only half of the work $A_{sn} = V\mu_{0}H_{c}^{2}$ is spent on creating the energy of the magnetic field $E_{m} = V\mu_{0}H_{c}^{2}/2$ in the normal state, and called the other half of the work as 'surplus work'. 

The opinion of W.H. Keesom and other physicists about the surplus work changed after the discovery of the Meissner effect. W.H. Keesom wrote in 1934: "{\it Till now we imagined that the surplus work served to deliver the Joule-heat developed by the persistent currents the metal getting resistance while passing to the non-supraconductive condition. As, however, the conception of Joule-heat can rather difficult be reconciled with reversibility we think now that there must be going on another process that absorbs energy}" \cite{Keesom1934}. The desire to think that the transition had become reversible was so great that W.H. Keesom made a false statement: "{\it So we admit that in passing the threshold value curve, what happens first is the penetrating of the magnetic field into the supraconductive material. By this process which is to be considered as reversible, the persistent currents are annihilated by induction}" \cite{Keesom1934}. Hirsch drew attention to the fallacy of the Keesom statement that magnetic induction annihilates the persistent currents rather than induces the Foucault currents: "{\it Keesom erroneously stated that the supercurrent stops by "induction", \cite{Keesom1934} but this is not so}" \cite{Hirsch2021JAP}. 

The illusion that the Meissner effect was able to make the irreversible transition reversible arose because the bulk superconductor returns to its initial state at the transition to the superconducting state in the magnetic field $H = H_{c}(T)$. But this return is observed not because the electromotive force has ceased to perform the work (10) when the magnetic flux increases from $\Phi = 0$ to $\Phi = \mu_{0}H_{c}\pi R^{2}$, but because the electromotive force performs the work of the opposite sign $A_{ns} = -A_{sn}$ when the magnetic flux decreases from $\Phi = \mu_{0}H_{c}\pi R^{2}$ to $\Phi = 0$ due to the Meissner effect. The transition at which the work of the opposite sign is performed cannot be considered a phase transition since the free energies must be equal 
$$F_{nH} = F_{sH} \eqno{(14)}$$ 
at the phase transition, and the work $A_{sn} = V\mu_{0}H_{c}^{2}$ performed on the system must increase its free energy $F = \Omega - ST$, according to the law of energy conservation $A_{sn} = \Delta \Omega $, if not all the work generates heat $A_{sn} > \Delta ST$. $\Omega $ is the sum of the system's internal energy. 

Nevertheless, after 1933, all physicists believed that superconducting transition of a bulk superconductor at $H = H_{c}(T)$ is the first-order phase transition. But in order to be able to think this way, it was necessary to forget either that the free energies should be equal (14) during the phase transition, or that the work changes the free energy. It seems incredible, but the authors of some books on superconductivity have really forgotten that work changes free energy \cite{Shoenberg1952,Lynton1962,Abrikosov1988,Schmidt1997}, while the authors of other books \cite{Tinkham,Ginzburg1946,Gennes1966} have forgotten that free energies must be equal at a phase transition. 

The equality of the free energies (14) cannot be deduced from the equalities (12) and (13), since without a magnetic field $H = 0$, the free energy of the superconducting state must be lower than the one of the normal state $F_{s0}(T_{1}) < F_{n0}(T_{1})$ at $T_{1} < T_{c}$. The authors \cite{Shoenberg1952,Lynton1962,Abrikosov1988,Schmidt1997}, who remembered that the free energies should be equal during the phase transition were forced to assert that the magnetic field increases the free energies of the superconducting state by an amount greater than that of the normal state, contrary to (12) and (13). But even between these authors there were disagreements. The authors \cite{Shoenberg1952,Lynton1962,Schmidt1997} postulate, contrary to the law of energy conservation, that the magnetic field increases the free energy not in the normal state
$$F_{nH}(T_{0}) = F_{n0}(T_{0})     \eqno{(15)}$$
but in the superconducting state
$$F_{sH}(T_{1}) = F_{s0}(T_{1}) + E_{m} = F_{s0}(T_{1}) + \frac{V\mu_{0}H^{2}}{2} \eqno{(16)}$$ 
A.A. Abrikosov \cite{Abrikosov1988} knew that the energy of the magnetic field in the volume of a superconductor is zero $VBH_{c}/2 = 0$, and was agreeing with (13). But he argued that the energy of the magnetic field does not increase, but decreases the free energy of the normal state $F_{nH}(T_{0}) = F_{n0}(T_{0}) - V\mu_{0}H^{2}/2$ \cite{Abrikosov1988}. This statement by Abrikosov contradicts the law of energy conservation. We have to do the work to create the energy of the magnetic field $E_{m} = V\mu_{0}H^{2}/2$. This work done on the system should increase its free energy, and A.A. Abrikosov argued that the work $E_{m} = V\mu_{0}H^{2}/2$ reduces the free energy \cite{Abrikosov1988}. 

The authors \cite{Tinkham,Ginzburg1946,Gennes1966}, unlike the authors \cite{Shoenberg1952,Lynton1962,Abrikosov1988,Schmidt1997}, did not contradict to the law of energy conservation. They agree with the equations (12) and (13) that the magnetic field increases the free energy of the normal state (12) and does not change the free energy of the superconducting state (13). They not only knew that the Faraday electromotive force performs the work (10) at the superconducting transition of a bulk superconductor in the magnetic field $H = H_{c}(T)$, but also calculated this work. P.G. de Gennes calculated the work (10) $A_{sn} = V\mu_{0}H_{c}^{2}$ during the transition from the superconducting state to the normal state \cite{Gennes1966}, and V.L. Ginzburg calculated the work of the opposite sign $A_{ns} = -A_{sn}$ during the transition from the normal state to the superconducting one \cite{Ginzburg1946}. 

V.L. Ginzburg wrote an equation for the change of free energy  
$$F_{sH} - F_{nH} = A_{ns} = -V\mu_{0}H_{c}^{2}  \eqno{(17)}$$  
during the transition from the normal to the superconducting one \cite{Ginzburg1946}. P.G. de Gennes and M. Tinkham wrote a similar equation 
$$F_{nH} - F_{sH} = A_{sn} = V\mu_{0}H_{c}^{2}  \eqno{(18)}$$ 
for the transition from the superconducting state to the normal one \cite{Tinkham,Gennes1966}. The free energy in accordance with (17) and (18) changes by an amount twice as large as in accordance with (11), since after the discovery of the Meissner effect, physicists stopped to think that the surplus work $A_{surp} = A_{sn} - E_{m} = V\mu_{0}H_{c}^{2}/2$ generates Joule heat. The surplus work allowed the authors \cite{Tinkham,Ginzburg1946,Gennes1966} to derive the same expression for the free energies difference between the normal and superconducting state at $H = 0$ and $T_{1} < T_{c}$ 
$$F_{n0} - F_{s0} = \frac{V\mu_{0}H_{c}^{2}}{2}    \eqno{(19)}$$ 
which the authors \cite{Shoenberg1952,Lynton1962,Abrikosov1988,Schmidt1997} deduced on the basis of the opposite statement that during the transition in the magnetic field $H = H_{c}(T)$ the free energy does not change (14) and no work is performed: according to \cite{Shoenberg1952,Lynton1962,Schmidt1997} and (14-16) $F_{nH} = F_{n0} = F_{sH} = F_{s0} + V\mu_{0}H_{c}^{2}/2$; according to A.A. Abrikosov \cite{Abrikosov1988} $F_{nH} = F_{n0} - V\mu_{0}H_{c}^{2}/2 = F_{sH} = F_{s0}$ and according to \cite{Tinkham,Ginzburg1946,Gennes1966} and (12), (13), (18) $F_{nH} - F_{sH} = F_{n0} + V\mu_{0}H_{c}^{2}/2 - F_{s0} = V\mu_{0}H_{c}^{2}$. 

It is amazing that famous physicists \cite{Tinkham,Ginzburg1946,Gennes1966} could forget that free energy cannot change during a phase transition. Even more surprising is the fact that V.L. Ginzburg \cite{Ginzburg1946} and P.G. de Gennes \cite{Gennes1966} calculated an entropy jump $S = -dF/dT$, which can be calculated only if the free energies are equal (14) at $H = H_{c}(T)$. According to (14-16) $dF_{nH}/dT = dF_{n0}/dT  = dF_{sH}/dT = dF_{s0}/dT + dE_{m}(H_{c})/dT = dF_{s0}/dT + V\mu_{0}H_{c}(dH_{c}/dT)$. Therefore, the jump in entropy between the normal state $S_{n} = -dF_{n0}/dT$ and superconducting state $S_{s} = -dF_{s0}/dT$ should be equal
$$S_{n} - S_{s} = V\mu_{0}H_{c}\frac{dH_{c}}{dT}  \eqno{(20)}$$  
The finite jump (20) cannot be deduced from (17) or (18), since the derivative $S = -dF/dT$ of the function (18), changing by a jump at $H = H_{c}(T)$ or $T_{c}(H)$ must be infinite: the free energy equals $F = F_{sH}$ at $T < T_{c}(H)$ and $F = F_{nH} = F_{sH} + 2E_{m}$ at $T > T_{c}(H)$ in accordance with (17) and (18). The finite jump (20) was deduced from (19) in the works \cite{Ginzburg1946,Gennes1966}. This deduction is mathematically incorrect, since the arguments of the left part (19) are $T = T_{1} < T_{c}$ and $H = 0$, whereas the arguments of the right part (19) are $T = T_{1} < T_{c}$ and $H_{c}(T)$. 

E.A. Lynton wrote in 1962 that the equation (19) "{\it is the basic equation of the thermodynamics of superconductors developed by Gorter and Casimir}" \cite{Lynton1962}. This thermodynamics, developed in \cite{Gorter1934}, played a crucial role in the emergence of the false notion about superconductivity and provoked the contradictions between the books on superconductivity, considered here. Equation (19) is the main equation, since all theories of superconductivity created after 1934 tried to explain why the difference $F_{n0} - F_{s0}$ in free energy between the normal $F_{n0}$ and superconducting $F_{s0}$ states at $T_{1} < T_{c}$ should be equal to the energy of the magnetic field in the normal state at $H = H_{c}(T_{1})$. Most physicists are confident that the BCS theory \cite{BCS1957} has successfully explained equation (19) derived by C.J. Gorter and H. Casimer \cite{Gorter1934}.  

But the equation (19) is questionable, since it was obtained on the basis of false statements and assumptions. C.J. Gorter and H. Casimer stated that "{\it the work, done by the current in the coil, which brings about $H$}" is equal to $dA = -HVdM$ \cite{Gorter1934}, where $VM = V(B - \mu_{0}H)$ is the total magnetic moment of the superconductor. The falsity of this statement provoked the contradiction of the authors \cite{Shoenberg1952,Lynton1962,Schmidt1997}, who followed the false statement of C.J. Gorter and H. Casimer \cite{Gorter1934}, with the authors \cite{Tinkham,Ginzburg1946,Gennes1966} who knew that the work (10) performed by the current in the coil is equal to $dA = HVdB$. The magnetic field increases the free energy of the superconducting state (16) rather than of the normal state (15), according to the books \cite{Shoenberg1952,Lynton1962,Schmidt1997} and contrary to the law of energy conservation since $M = B - \mu_{0}H = \mu_{0}H - \mu_{0}H = 0$ in the normal state and $M = B - \mu_{0}H = - \mu_{0}H$ in the superconducting state. The authors of the books \cite{Tinkham,Ginzburg1946,Gennes1966} deduced the opposite expressions (12) and (13), which do not contradict to the law of energy conservation, since $B = 0$ in the superconducting state and $B = \mu_{0}H$ in the normal state. 

The authors \cite{Shoenberg1952,Lynton1962,Schmidt1997} did not follow in everything C.J. Gorter and H. Casimer. According to the article \cite{Gorter1934} the work $A_{sn} = VH_{c}\Delta M = -V\mu_{0}H_{c}^{2}$ should be performed due to the change in the magnetic moment from $VM = V(B - \mu_{0}H_{c}) = - V\mu_{0}H_{c}$ to $VM = 0$ during the transition of a bulk superconductor to the normal state in the critical magnetic field $H = H_{c}(T)$, see (11) in \cite{Gorter1934}. To derive equation (19) C.J. Gorter and H. Casimer "{\it introduce the assumption, that the second law of thermodynamics applies also to the transition at}" $H = H_{c}(T)$ \cite{Gorter1934}. This assumption cannot be true because of the violation of this law in accordance with equalities (11) and (13) in the article \cite{Gorter1934}. The heat of the transition $Q_{2}$ at $H = H_{c}(T)$ is equal to the amount of work $Q_{2} = A_{sn} = VH_{c}\Delta M = -V\mu_{0}H_{c}^{2}$ during this transition in accordance with (11) of \cite{Gorter1934}. 

The work, calculated by C.J. Gorter and H. Casimer, has the same amount, but the opposite sign, compared to the work calculated by the authors \cite{Tinkham,Ginzburg1946,Gennes1966}, since $dA = -VHdM$ rather than $dA = VHdB$ according to the article \cite{Gorter1934}. The authors \cite{Tinkham,Ginzburg1946,Gennes1966} were sure that all, and not half, of the positive work $A_{sn} = V\mu_{0}H_{c}^{2}$ increases (18) the free energy $F = \Omega - ST$, since if half of the work (the surplus work) $A_{surp} = A_{sn} - E_{m} = V\mu_{0}H^{2}/2$ increases heat $Q = ST$, as physicists thought before 1933, then half of the negative work $A_{ns} = -V\mu_{0}H_{c}^{2}$ done during the transition from the normal state to superconducting one due to the Meissner effect \cite{Ginzburg1946} should be taken from heat, contrary to the second law of thermodynamics. C.J. Gorter and H. Casimer  \cite{Gorter1934} claimed that all the negative work $A_{sn} = -V\mu_{0}H_{c}^{2}$, that they calculated, is taken from heat, contrary to the second law of thermodynamics.

This absurd claim that the second law of thermodynamics can be violated even without the Meissner effect is a consequence of the desire of C.J. Gorter and H. Casimer \cite{Gorter1934} to prove that the Meissner effect does not contradict the second law of thermodynamics. The centuries-old faith, or superstition, about the impossibility of a perpetual motion machine has provoked a variety of opinions even about the work performed by Faraday's electromotive force during the transition of a bulk superconductor to the normal state in the magnetic field $H = H_{c}(T)$: this work is equal $A_{sn} = V\mu_{0}H_{c}^{2}$ according to W.H. Keesom \cite{Keesom1934} and the authors \cite{Tinkham,Ginzburg1946,Gennes1966}; $A_{sn} = 0$ according to the books \cite{Shoenberg1952,Lynton1962,Schmidt1997} and $A_{sn} = -V\mu_{0}H_{c}^{2}$ according to C.J. Gorter and H. Casimer \cite{Gorter1934}. 

This story with the Meissner effect shows that not only superstition, but also chance can play a decisive role in science. C.J. Gorter and H. Casimer, in order to derive the basic equation of their thermodynamics of superconductors (19), assumed that the total change in entropy in the Gorter cycle equals zero in accordance with the requirement of the second law of thermodynamics \cite{Gorter1934}. They would have to understand that this assumption cannot be correct if they calculated the total work performed in the Gorter cycle, which is equal to the surplus work $A_{surp} = A_{sn} - E_{m} = V\mu_{0}H^{2}/2$. This work increases either free energy $F = \Omega - ST$ or heat $Q = ST$. By definition, free energy cannot change in any closed cycle. 

Thus, the heat should increase by the amount of the surplus work $A_{surp} = V\mu_{0}H^{2}/2$ in each Gorter cycle, contrary to the assumption of C.J. Gorter and H. Casimer \cite{Gorter1934}. This conclusion means that the Meissner effect refutes the second law of thermodynamics, since in each cycle opposite to the Gorter cycle, the negative surplus work $A_{surp} = -V\mu_{0}H^{2}/2$ must be performed: 1) applying a magnetic field $H$ in the normal state at a temperature $T_{0} > T_{c}$; 2) cooling from $T_{0} > T_{c}$ to a temperature $T < T_{c}$ below the transition into superconducting state at $H = H_{c}$; 3) switching off the magnetic field $H = 0$ in the superconducting state at $T < T_{c}$; 4) temperature increase from $T < T_{c}$ to $T_{0} > T_{c}$ at zero external field $H = 0$. If C.J. Gorter and H. Casimer, W.H. Keesom, or someone else calculated the work $A_{surp} = V\mu_{0}H^{2}/2$ performed in the Gorter cycle and the cycle opposite to the Gorter cycle, they would have understood that violations of the second law of thermodynamics cannot be avoided. 

\section{Conclusion}
The contradictions between the authors of the books, which are highlighted in the previous section, are surprising. How could it happen that for many years no one paid attention to this contradiction? V.L. Ginzburg wrote in 1946 the expression (17) for the change in the free energy during the transition at $H = H_{c}$ based on the fact that "{\it As is known from thermodynamics, the difference in free energies after the transition and before it $F_{sH} - F_{nH}$ is equal to the work done on the body}" \cite{Ginzburg1946}. Unlike V.L. Ginzburg, A.A. Abrikosov stated in 1988: "{\it The condition of a superconducting transition is equality of free energies}" \cite{Abrikosov1988}. How could A.A. Abrikosov not know what V.L. Ginzburg wrote about in 1946?  

It should be said that A.A. Abrikosov used a trick to substantiate his statement. The difference in free energies after the transition and before it should be equal to the work $A_{sn} = V\mu_{0}H_{c}^{2}$ done on the superconductor, according to the expression (P.3.5) $dF = dF_{0} + HdB$ in  Appendix 3 to Abrikosov's book \cite{Abrikosov1988}. M. Tinkham called this free energy Helmholtz free energy \cite{Tinkham}. A.A. Abrikosov, based on the statement that the use of this free energy in the case of a superconductor is inconvenient, used free energy $G = F - VHB$, which M. Tinkham called Gibbs free energy \cite{Tinkham}. The Gibbs free energy is more convenient only because the change in the value $VH_{c}B$ is equal to the work $A_{sn} = V\mu_{0}H_{c}^{2}$ performed when the magnetic flux density changes from $B = 0$ to $B = \mu_{0}H_{c}$ in the volume $V$ of the superconductor. Therefore, it is possible to write down the equality of free energies $G_{nH} = G_{sH}$ that must be present at any phase transition, despite of the work $A_{sn} = V\mu_{0}H_{c}^{2}$ performed at the superconducting transition. 

The trick with the Gibbs free energy cannot eliminate the work $A_{sn} = V\mu_{0}H_{c}^{2}$ done during the superconducting transition at $H = H_{c}$. This trick can only justify the illusion that this transition is a phase transition. The authors \cite{Shoenberg1952,Lynton1962} used the Gibbs free energy to justify the equalities (15) and (16). This justification cannot be correct since both the Helmholtz free energy $F$ and the Gibbs free energy $G = F - VHB$ cannot increase with the magnetic field in the superconducting state in which the magnetic flux density is zero $B = 0$. A.A. Abrikosov is more right in saying \cite{Abrikosov1988} that the magnetic field does not change the Gibbs free energy $G = F - VHB$ in the superconducting state and decreases the free energy in the normal state since $G_{nH} = F_{nH} - VHB = F_{n0} + V\mu_{0}H^{2}/2 - V\mu_{0}H^{2} = G_{n0} - V\mu_{0}H^{2}/2$. The equalities (15) and (16) written in \cite{Shoenberg1952,Lynton1962,Schmidt1997} can be deduced only from the incorrect statement of C.J. Gorter and H. Casimer that "{\it the work, done by the current in the coil, which brings about $H$}" is $dA = -HVdM$ \cite{Gorter1934}. 

Hirsch drew attention to the Meissner effect puzzle because of the confidence that his theory of hole superconductivity could solve this puzzle \cite{Hirsch10Meis}. But Hirsch's theory cannot replace the conventional theory of superconductivity \cite{GL1950}, which explains, as a consequence of quantization (1), not only the Meissner effect \cite{Meissner1933}, the magnetic flux quantization \cite{fluxquan1961a,fluxquan1961b} and quantum oscillation of the persistent currents \cite{PCJETP07,JETP07J}, but also the Abrikosov state \cite{Abrikosov}. Abrikosov vortices are singularities of the GL wave function $\Psi _{GL} =|\Psi _{GL}|\exp{i\varphi }$ that allow the magnetic flux to penetrate deep into the superconductor according to the GL expression (5), due to the fact that the change of the phase of the GL wave function around each vortex is not zero, but $2\pi$, $\oint_{l}dl \nabla \varphi = 2\pi $ and $\oint_{l}dl \nabla \varphi = n2\pi $ around $n$ Abrikosov vortices. The magnetic flux can be equal to $\Phi = n\Phi_{0}$ inside a macroscopic closed path $l$ in accordance with (5), where $n$ is the number of the Abrikosov vortices inside $l$. 

But Hirsch is right in the sense that quantization and violation of the correspondence principle can explain the macroscopic change in the angular momentum of mobile charge carriers, but not of the crystalline lattice of superconductor ions. This is not the only puzzle that seems unsolvable. Another such puzzle is discussed in publications \cite{PLA2012T,Nikulov2016}. The GL theory \cite{GL1950} predicts that a potential difference with a constant component $V_{dc} \propto \overline{I_{p}}$ should be observed when a segment of a ring with the persistent current (6) is switched between superconducting and normal state with a sufficiently high frequency $f_{sw}$ \cite{LTP1998}. Observations \cite{PCJETP07,Letter2003,Physica2019,PLA2012Ex,APL2016,PLA2017,Letter2007} of the dc voltage $V_{dc}(\Phi /\Phi _{0}) \propto \overline{I_{p}}(\Phi /\Phi _{0})$ corroborate this prediction. But there is unsolved puzzle in this prediction. The GL theory \cite{GL1950} predicts that Cooper pairs in a ring segment $l_{B}$ should start moving due to quantization (3), when the GL wave function $\Psi _{GL} =|\Psi _{GL}|\exp{i\varphi }$ closes in a remote segment $l_{A}$ during its transition to the superconducting state \cite{PLA2012T,Nikulov2016}. But no theory can say how quickly the velocity of Cooper pairs on the segment $l_{B}$ will become non-zero after the transition of the segment $l_{A}$ and how the time between these events may depend on the distance between the segments $l_{A}$ and $l_{B}$.  

Hirsch has drawn attention \cite{HirschEPL,HirschIJMP,HirschPhys} to the internal inconsistency of the conventional theory of superconductivity \cite{BCS1957}, because of his confidence that his theory can solve the Keesom task \cite{Keesom1934}: to explain how the persistent currents can be annihilated before the material gets resistance. He did not take into account that this task lost its meaning \cite{PhysicaC2021} after experimental evidence obtained firstly in 1962 \cite{LP1962} that the persistent currents are not annihilated even after the material gets resistance. Moreover experiments \cite{PCJETP07,Letter2003,Physica2019,PLA2012Ex,APL2016,PLA2017,Letter2007} give evidence that the persistent current $\overline{I_{p}} \propto V_{dc}$ can flow against the dc electric field $E = -\nabla V_{dc} $ in one of the ring segments \cite{Physica2019,PLA2012T}. The GL theory \cite{GL1950} can explain even this paradoxical phenomenon as a consequence of the emergence of the persistent current (6) because of quantization (5) in the superconducting state and its disappearance in the normal state due to the dissipation of its energy into Joule heat. 

It should be emphasized that not only this explanation, but also observations the persistent currents $\overline{I_{p}} \neq 0$ at non-zero resistance $\overline{R} > 0$ \cite{Letter2007,Science2007,LP1962,LP2010Nature,LP2010PRB,LP1990PRB} and the DC power $\overline{I_{p}}V_{dc}$ \cite{Physica2019}, as well as the Meissner effect \cite{Entropy2022}, contradict the second law of thermodynamics \cite{PhysicaC2021}. Physicists have not noticed this contradiction for many years because of their confidence, especially characteristic of theorists, that they can explain all observed phenomena without going beyond their beliefs. This confidence and insufficiently critical attitude to successful theories leads to the degradation of physical thinking. The Comment \cite{EPL2021} and the article \cite{FoundPhys2023} draw attention on examples of such degradation. 
              
This work was made in the framework of State Task No 075-00296-24-01. The author is grateful to Prof. Jorge Hirsch for the useful discussions. These discussions made it possible, in particular, to notice contradictions in books on superconductivity that no one had noticed for many years.

\end{document}